\documentclass[twocolumn,english,aps,prl,floatfix,]{revtex4}
\usepackage[T1]{fontenc}
\usepackage[latin1]{inputenc}
\usepackage{graphicx}
\usepackage{epstopdf}
\DeclareGraphicsExtensions{.eps}
\usepackage{amsmath}
\usepackage{tabularx}
\makeatletter

\usepackage{babel}

\usepackage{babel}
\makeatother

\begin{document}

\title{Quantum Monte Carlo study of the itinerant-localized model of strongly correlated electrons: Spin-spin
correlation functions}

\author {Ilya Ivantsov$^{1,3}$,
Alvaro Ferraz$^{2}$, and Evgenii Kochetov$^{3}$}

\affiliation{$^{1}$L.V.Kyrensky Institute of Physics, Siberian Branch of Russian Academy of Sciences, Krasnoyarsk, Russia}
\affiliation{$^{2}$International Institute of Physics - UFRN,
Department of Experimental and Theoretical Physics - UFRN, Natal, Brazil,}
\affiliation{$^{3}$Bogoliubov Laboratory of Theoretical Physics, Joint
Institute for Nuclear Research, Dubna, Russia}


\begin{abstract}
We perform quantum Monte Carlo simulations of the itinerant-localized periodic Kondo-Heisenberg model for the underdoped cuprates
to calculate the associated spin correlation functions.
The strong electron correlations are shown to play a key role in
the abrupt destruction of the quasi long-range antiferromagnetic order
in the lightly doped regime.
\end{abstract}

\maketitle

\section{Introduction}
The aim of this Letter is to explore the mechanism underlying the abrupt suppression
of the long range antiferromagnetic (AF) order  observed in the lightly hole-doped cuprates.
As is well known the $2d$ undoped quantum AF exhibits at zero temperature the AF long range order (LRO) that is completely destroyed
by a surprisingly low doping. It is very reasonable to assume that the strong electron correlations are at work in this case.
Technically, the lightly doped regime is pretty hard to address because, precisely under this condition, the constraint
of no double electron occupancy (NDO) is fully at work. This implies that, due to the strong on-site Coulomb repulsion, two lattice electrons cannot hop onto one and the same lattice site regardless of their spin projection. Such a local restriction on a structure of the Hilbert space is very hard to implement analytically in a reliable and controlled manner.
Alternative slave-particle mean-field theories that treat the local NDO constraint
only globally predict a nonphysically large value of the critical doping.\cite{lee}

The NDO constraint  drives the theory into a strong-coupling regime, which calls for
proper technical tools.
Some progress can be achieved by employing
the earlier established mapping of the $t-J$ model of strongly correlated electrons onto the Kondo-Heisenberg model at
a dominantly large Kondo coupling \cite{pfk}. Being a slave-particle theory, such an approach possesses however
a few important advantages over the conventional slave-particle theories.

First of all, the strength of electron correlations is now
encoded into a single global parameter - a Kondo coupling. Varying its magnitude enables us to get important insights
as to in what way the strong electron correlations affect the underlying physics.
In particular we show that the local NDO constraint is responsible for a
rapid destruction of the  AF quasi LRO (QLRO) with doping. If the NDO constraint is ignored the QLRO is restored.
The critical hole concentration at which the AF QLRO disappears acquires a reasonably low
value.

Additionally, the proposed spin-dopon theory explicitly takes into account the dual nature of the
constrained lattice electrons.
In the underdoped cuprates, one striking feature is the simultaneous
localized and itinerant nature of the lattice electrons.
Such a duality appears as an explicit manifestation of the local Mott physics
and is shown to be a direct consequence of the local NDO constraint.

Moreover, the itinerant-localized model provides a convenient new set of coordinates well suited for  numerical simulations.
Specifically,
one can vary the strength of the electron correlations by simply varying a single global parameter - the (Kondo)
coupling between the itinerant and localized electrons.
In particular, classical Monte Carlo simulations for large clusters were successfully used in studying the electron spin correlations
in the full Ising version of the $2d$ $t-J$ model in the spin-dopon representation.\cite{marcin} It was shown that the AF LRO disappears already at the doping of the order of a few percent. It was also demonstrated that the NDO constraint is responsible for the
smearing out of the magnetic order.
However, these results were obtained within a simplified model with the transverse components of the on-site electron spin being
self-consistently neglected.

In the present paper,
we apply the {\it quantum} Monte Carlo (QMC) simulations to explore the quantum spin dynamics of the
underdoped cuprates within the standard $SU(2)$ invariant $2d$ $t-J$ model.
We intend to explore the issue as to whether or not
the NDO constraint still plays a dominant role in the disruption of the magnetic order
in the lightly doped regime. One should however keep in mind that
the QMC method restricts ourselves to deal with finite temperatures and finite lattice clusters.
As a result, this approach cannot capture a true LRO in $2d$. Since the AF correlation length remains finite,
we consider a finite-size system away from the critical point. A full theory of such systems is not available yet.
However, at sufficiently low doping, the correlation length is much larger than a characteristic cluster size. This manifest itself as a QLRO. What is important is that the QMC method enables us to observe a rapid destruction of the QLRO with increasing doping and the formation of the short range
order (SRO) instead. We explicitly demonstrate that the local NDO constraint plays a dominant role in destroying the magnetic order at finite doping in the standard $t-J$ model.

\section{Model}
To start with, let us briefly review the Kondo-Heisenberg-model approach to strongly correlated electron systems.
The canonical $t-J$ model Hamiltonian of strongly correlated electrons reads
\begin{equation}
H_{t-J}=-\sum_{ij\sigma} t_{ij} \tilde{c}_{i\sigma}^{\dagger}
\tilde{c}_{j\sigma}+ J\sum_{ij} (\vec Q_i \cdot \vec Q_j -
\frac{1}{4}\tilde{n}_i\tilde{n}_j),
\label{1.1}\end{equation}
where $\tilde{c}_{i\sigma}=c_{i\sigma}(1-n_{i,-\sigma})$ is the projected electron operator, $\vec
Q_i=\sum_{\sigma,\sigma'}\tilde{c}_{i\sigma}^{\dagger}\vec\tau_{\sigma\sigma'}\tilde{c}_{i\sigma'},
$ is the electron spin operator, $\tilde
n_i=n_{i\uparrow}+n_{i\downarrow}-2n_{i\uparrow}n_{i\downarrow}$ and $\vec\tau$ is the Pauli vector, $\vec\tau^2=3/4$.
In the underdoped cuprates, one striking feature is a simultaneous display of both localized and itinerant nature of the lattice electrons, $\tilde{c}_{i\sigma}$.
To include these both aspects of the constrained electrons into consideration, on equal footing,
Ribeiro and Wen proposed a slave-particle spin-dopon representation of the projected electron operators
in the enlarged Hilbert space \cite{wen},
\begin{eqnarray}
\tilde c_{i}^{\dagger}
=\frac{1}{\sqrt{2}}(\frac{1}{2}-2\vec S_i\vec\tau)\tilde d_i.
\label{1.2}\end{eqnarray}
In this framework, the localized electron is represented by the lattice spin $\vec S\in su(2)$
whereas  the doped hole (dopon) is described by the projected hole operator,
$\tilde{d}_{i\sigma}=d_{i\sigma}(1-n^d_{i-\sigma})$.
Here $\tilde c^{\dagger}=(\tilde c^{\dagger}_{\uparrow},\tilde c^{\dagger}_{\downarrow})^{t}$ and
$\tilde d=(\tilde d_{\uparrow}, \tilde d_{\downarrow})^{t}$.

The physical content of the spin-dopon representation (\ref{1.2}) can be clarified as follows.
First of all, we represent the  spin degrees of freedom in terms of
chargeless fermions (spinons), $f_{\sigma}:$
$$\vec
S=\sum_{\sigma,\sigma}f_{\sigma'}^{\dagger}\vec\tau_{\sigma'\sigma}f_{\sigma}, \quad
\sum_{\sigma}f^{\dagger}_{\sigma}f_{\sigma}=1.$$
Following this, we introduce
the operator \cite{pfk} $$ D=\frac{f_{\uparrow}\tilde d_{\downarrow}-f_{\downarrow}\tilde d_{\uparrow}}{\sqrt{2}}$$
which destroys the on-site  spin-dopon singlet state (holon).
The physical electron operator (\ref{1.2}) then reduces to the spinon-holon decomposition:
\begin{equation}
\tilde c_{\sigma}^{\dagger}=f^{\dagger}_{\sigma} D.
\label{sb}\end{equation}
This equation appears as a  slave-boson representation of the constrained electron operator in terms of the itinerant and localized degrees of freedom with the boson being a composite state.
The itinerant boson (holon) appears as a charged spinon-dopon singlet and it corresponds to a
hopping vacancy. The localized lattice spin is represented by a chargless spinon state that transforms as an $SU(2)$ spinor.

The physical on-site Hilbert space is a $3d$ one that comprises  spin-up, spin-down states, and a vacancy.
In terms of the projected electron operators, the NDO constraint to single out the physical  Hilbert space takes the form
\begin{equation}
\sum_{\sigma}(\tilde c_{i\sigma}^{\dagger}\tilde c_{i\sigma})+\tilde c_{i\sigma}\tilde c_{i\sigma}^{\dagger}=1.
\label{a} \end{equation}
Only under  this condition are the projected electron operators
isomorphic to the Hubbard operators.
Within the spin-dopon representation, the NDO reduces to a Kondo-type interaction constraint \cite{pfk},
\begin{equation}
\vec{S_i} \cdot
\vec{s_i}+\frac{3}{4}(\tilde{d}_{i\uparrow}^{\dagger}\tilde{d}_{i\uparrow}+
\tilde{d}_{i\downarrow}^{\dagger}\tilde{d}_{i\downarrow})=0,
\label{cnstr}\end{equation}
with $\vec
s_i=\sum_{\sigma',\sigma}\tilde{d}_{i\sigma'}^{\dagger}\vec\tau_{\sigma'\sigma}\tilde{d}_{i\sigma}
$  being the dopon spin operator.
Equivalently, Eq.(\ref{cnstr}) can be written in the form
$ D_i^{\dagger}D_i=\tilde n_i^d.$

At strong coupling $(\lambda\gg t)$, the original $t-J$ model (\ref{1.1}) is shown to be equivalent
to the lattice Kondo-Heisenberg-type  model \cite{pfk}:
\begin{eqnarray}
H_{t-J}&=& \sum_{ij\sigma} 2t_{ij}
{d}_{i\sigma}^{\dagger} {d}_{j\sigma}
+ J\sum_{ij} \vec S_i(1-n_i^d) \cdot \vec S_j(1-n_j^d)\nonumber\\
&+& \lambda
\sum_i(\vec{S_i} \cdot
\vec{s_i}+\frac{3}{4}{n}^d_i),\quad \lambda\to +\infty,
\label{2.7}
\end{eqnarray}
where we have dropped the "tilde" sign of the dopon operators, as it becomes irrelevant due to the NDO constraint.
The unphysical doubly occupied electron states are separated from the physical
sector by an energy gap $\sim\lambda$.
In the $\lambda\to +\infty$ limit, i.e. in the limit in which $\lambda$ is much larger than any other existing energy
scale in the problem, those states
are  automatically excluded from the Hilbert space.
In spite of the global character of the parameter $\lambda$, it  enforces the NDO constraint locally due to the fact that
the on-site physical Hilbert subspace corresponds to zero eigenvalues of the constraint, whereas the nonphysical subspace
is spanned by the eigenvectors with strictly positive eigenvalues. In $1d$, Eq.(\ref{2.7}) reproduces
the well-known exact results for the $t-J$ model \cite{fk} (see also Appendix).

Close to half filling, where the density of doped holes
is small $\delta:=\langle n^d_i\rangle \ll 1$, one can make the change $J\to \tilde J=J(1-\delta)^2.$
The spin-dopon representation of the $t-J$ Hamiltonian for
the underdoped cuprates then reduces to the Kondo-Heisenberg lattice model at a dominantly large Kondo coupling \cite{pfk},
\begin{equation} H_{t-J} = \sum_{ij\sigma}
t^{eff}_{ij}{d}_{i\sigma}^{\dagger} {d}_{j\sigma}+
\tilde{J}\sum_{ij} (\vec S_i \cdot \vec S_j - \frac{1}{4}) +\lambda
\sum_{i}\vec{S_i} \cdot \vec{s_i},
\label{04}\end{equation}
where $t^{eff}_{ij}=2t_{ij}+(3\lambda/4-\mu) \delta_{ij}$ and $\lambda\gg t,J$.

In the spin-dopon representation (\ref{04}), the on-site Hilbert space is spanned by the vectors $|\sigma a\rangle$, with $\sigma=\uparrow,\downarrow$ labeling the lattice spin projection and $a=0,\uparrow,\downarrow$ labeling dopon state.
Explicitly they are numerated by an integer $p=1,2,..6$ as given in Table \ref{tab:1}.
\begin{table}[h]
\caption{The basis states.}
\label{tab:1}
\begin{center}
\begin{tabular}{|c|c|c|c|c|c|}
\hline
$|1\rangle$ & $|2\rangle$ & $|3\rangle$ & $|4\rangle$ & $|5\rangle$ & $|6\rangle$ \\
\hline
~~$|\uparrow~\uparrow\rangle$~~ & ~~$|\uparrow 0\rangle$~ & ~$|\uparrow~\downarrow\rangle$~~ & ~~$|\downarrow~\uparrow\rangle$~~ & ~~$|\downarrow 0\rangle$~~ & ~~$|\downarrow~\downarrow\rangle$~~ \\
\hline
\end{tabular}
\end{center}
\end{table}

Any on-site operator $A$ can then be identically written in the form
$A_i=\sum_{pq}\langle p|A_i|q\rangle X^{pq}_i$, where $X^{pq}:=|p\rangle\langle q|.$

Since we are interested in the large $\lambda$ limit, it seems appropriate to separate the Hamiltonian in the following way:
$H_{t-J}=H_{\lambda}+H_{z}+H_{int}$, where
\begin{equation}
\begin{gathered}
H_\lambda=\lambda \sum_{i} \sum_{pq}(\frac{3}{4}\langle p|n_i^d|q\rangle + \langle p|\vec S_i\cdot \vec s_i|q\rangle)X^{pq}_i,\\
H_z=\tilde{J}\sum_{ij}\sum_{pq}\langle p|S_i^z|q\rangle\langle m|S_j^z|n\rangle X^{pq}_i X^{mn}_j,\\
H_{int}=\sum_{ij\sigma}\sum_{pqmn}2t_{ij}\langle p|{d}_{i\sigma}^{\dagger}|q\rangle\langle m|{d}_{j\sigma}|n\rangle X^{pq}_i X^{mn}_j +\\
+\frac{\tilde{J}}{2}\sum_{ij}\sum_{pqmn}\langle p|S_i^+|q\rangle\langle m|S_j^-|n\rangle X^{pq}_i X^{mn}_j + h.c.
\end{gathered}
\end{equation}
In this basis $H_\lambda$ takes on a non diagonal form:
\begin{equation}
H_\lambda=\lambda\sum_i (X^{11}_i+X^{66}_i+\frac{1}{2}(X^{33}_i+X^{34}_i+X^{43}_i+X^{44}_i)).
\end{equation}
This form is inconvenient for numerical purposes, however. In the large $\lambda$ limit,
the probability of the updating procedure involving $\lambda$ becomes much higher than the others.
This leads to a crucial slowdown of the calculations. To get around this problem,
it is more convenient to go over to the basis constructed out of the eigenstates of $H_{\lambda}$ as given in Table \ref{tab:2}.
\begin{table}[h]
\caption{The new basis states.}
\label{tab:2}
\begin{center}
\begin{tabular}{|c|c|c|c|c|c|}
\hline
$|1\rangle$ & $|2\rangle$ & $|3\rangle$ & $|4\rangle$ & $|5\rangle$ & $|6\rangle$ \\
\hline
~~$|\uparrow~\uparrow\rangle$~~ & ~~$|\uparrow 0\rangle$~~ & ~~$\frac{|\uparrow\downarrow\rangle-|\downarrow\uparrow\rangle}{\sqrt{2}}$~ & ~$\frac{|\uparrow\downarrow\rangle+|\downarrow\uparrow\rangle}{\sqrt{2}}$~ & ~$|\downarrow 0\rangle$~ & ~$|\downarrow~\downarrow\rangle$~~ \\
\hline
\end{tabular}
\end{center}
\label{tab2}
\end{table}
In this case, the $H_\lambda$ becomes diagonal in the $(p,q)$ representation:
\begin{equation}
H_\lambda=\lambda\sum_i (X^{11}_i+X^{66}_i+X^{44}_i),
\label{10}
\end{equation}
where the unphysical spin triplet states $|p\rangle,\, p=1,4,6,$ enter with eigenvalue $\lambda$.
The physical vectors $|p\rangle,\, p=2,3,5,$ that describe the vacancies and lattice spins correspond to zero eigenvalues of $H_{\lambda}$.
Due to the fact that the statistical weights of configurations with states $|p\rangle$, $p=1,4,6$, are proportional to $e^{-\beta\lambda}$, we can exclude these states from calculation, provided $\lambda$ is large enough.
From now on all the states denoted by $|p\rangle$ correspond to those from Table II.

\section{Method}

In our calculations, we use the Contnuous Time WorldLine (CTWL) QMC method.
Following \cite{ctwl} the algorithm is modified by adding "worms" in the representation of
the $X$-operators, which corresponds to the addition of a fictitious term to the Hamiltonian:
\begin{equation}
H_{\nu}=\sum_{ipq}\nu_{pq}(X_i^{mn}+X_i^{nm}),
\label{Hnu}
\end{equation}
where $\nu_{pq}$ is a set of fictitious amplitudes satisfying $\nu_{pp}=0$ and $\nu_{pq}=\nu_{qp}$ chosen to improve a convergence.
This terms are included in the non diagonal part corresponding to the existence of a worm in the configuration.
Since all measurements occur in the absence of the worms, they do not contribute to the final result.
This update allows us to make calculations more effective by adding the fictitious configurations to the true ones.
In particular, one is able to run calculation in the grand canonical ensemble keeping at the same time a total number of particles under control.

QMC method is based on the representation of the partition function in the interaction picture\cite{qmc}:
\begin{equation}
e^{-\beta H_{t-J}} =e^{-\beta H_0} T_{\tau}(exp(-\int_0^\beta H_1(\tau)d\tau)),
\label{QMC}
\end{equation}
where $T_\tau$ denotes the $\tau$-ordering operator, and
\begin{eqnarray}
H_{t-J}&=&H_0+H_1,\nonumber\\
H_0&=&H_\lambda+H_z^{diag},\\
H_1&=&H_{int}+H_\nu+H_z^{nondiag}.\nonumber
\end{eqnarray}
The partition function expansion takes the form:
\begin{eqnarray}
Z &=& Sp(e^{-\beta H_0}(1-\int_0^{\beta}H_1(\tau)d\tau+\nonumber\\
&+&\int_0^{\beta}\int_0^{\tau_1}H_1(\tau_1)H_1(\tau_2)d\tau_1 d\tau_2) -...),
\label{PF}
\end{eqnarray}
where $$H_1(\tau):=e^{-\tau H_0}H_1e^{\tau H_0}.$$
The representation (\ref{QMC}-\ref{PF})) allows us to consider the cases of large  and small $\lambda$ on equal footing.
In either case, $H_0$ represents a leading contribution to the partition function. In particular, a quasi long-range order
restores at  small $\lambda$ in which case the $S^zS^z$ interaction term in $H_0$ becomes of the major importance.

\begin{figure}
\begin{minipage}[h]{1\linewidth}
\center{\includegraphics[width=1\linewidth]{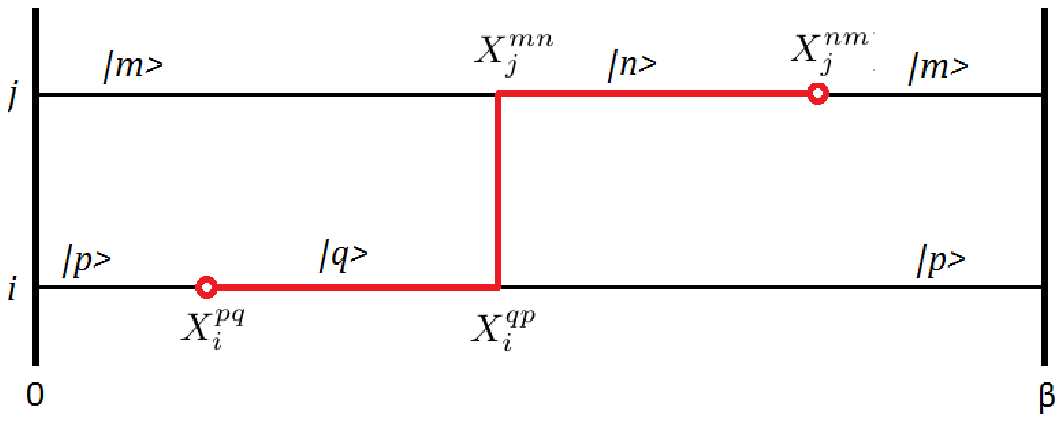}} a) \\
\end{minipage}
\hfill
\begin{minipage}[h]{1\linewidth}
\center{\includegraphics[width=1\linewidth]{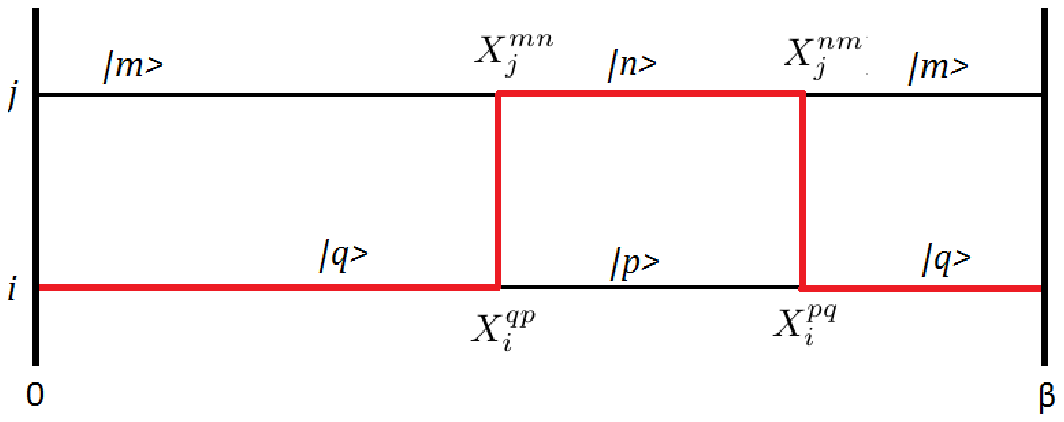}} b) \\
\end{minipage}
\caption{The fragments of typical configurations. Panels show configurations with (a) and without (b) a worm. In the panel (a),
the red color indicates the worldline of the worm; in the panel (b), the red color indicates the loop formed by the worm closing.}
\label{ris:qmc}
\end{figure}

In CTWL method, the expansion of the partition function comes in the form of the worldlines states in the imaginary time. It is convenient to  rewrite the Hamiltonian in a form  suitable for this method:
\begin{eqnarray}
H_{t-J}=\sum_{i,p}E_p X_i^{pp}+\sum_{ij,pq}V_{ij}^{pq}X_i^{pp}X_j^{qq}+\nonumber\\
+\sum_{ij,pqrs}T_{ij}^{pqrs}X_i^{pq}X_j^{rs}+\sum_{i,pq}\nu_{pq}(X_i^{pq}+X_i^{qp}),
\label{PF1}
\end{eqnarray}
where $E_p$ are the eigenvalues of the one-site part of the Hamiltonian (\ref{10})) whereas
the $V_{ij}^{pq}$ are the energies of the diagonal interaction between the states $|p\rangle$ and $|q\rangle$ on sites $i$ and $j$, respectively.
In this form, to each segment of the horizontal line that lies between $\tau_1$ and $\tau_2$ in state $|p\rangle$, there corresponds the multiplier $e^{-\beta (E_p+V) d\tau}$, where $V$ is the average energy of the diagonal interaction over all neighboring segments and $d\tau=\tau_2-\tau_1$ is the imaginary time interval of the segment.

To each kink (a segment of the vertical line in Fig.\ref{ris:qmc}) between sites $i$ and $j$ there corresponds the multiplier $T_{ij}^{pqmn}X_i^{pq}X_j^{mn}$, where $T_{ij}^{pqmn}$ are the energies of nondiagonal interaction that corresponds to the state change. Also, with respect to the worm algorithm, to each discontinuities there corresponds the multiplier $\nu_{pq}X_i^{pq}$ that represents the worm boundary, where $\nu_{pq}$ is the above introduced fictitious energy.

Fig.\ref{ris:qmc} shows the fragments of a typical configuration occurring during the simulation.
The update of the configurations occurs according to the Metropolis algorithm\cite{metropolis} involving finite number of updating procedures.
Those updating procedures are described in detail in \cite{ctwl}. However, to increase the convergence
as well as to ensure the ergodicity of the algorithm, it must be also accompanied  by certain additional
prescriptions \cite{troyer}.

The observables are measured in the following way:
\begin{equation}
\langle A\rangle:=\frac{\sum_{MC}\langle p|A e^{-\beta H}|p\rangle}{\sum_{MC}1}
\label{Obs}
\end{equation}
where $A$ is some operator. Unfortunately in the fermion system we are faced with \textit{the sign problem}.
This problem is connected with the appearance of the negative statistical weights during the calculation.

\begin{equation}
\langle A\rangle=\frac{\sum_{MC}\langle p|A e^{-\beta H}|p\rangle sign(W)}{\sum_{MC}sign(W)}
\label{Sign}
\end{equation}
As a result, the errors increase exponentially with decreasing temperature which rules out an acceptable accuracy at low temperatures.
Finally, the adopted algorithm goes through the following steps:

$\textbf{(i)}$ an initial configuration is generated.
In fact, the initial configuration selection has no impact on the final result. All possible
impacts of this choice are eliminated by thermalization.

$\textbf{(ii)}$ possible updating procedures are chosen randomly.
The probabilities of the procedures are not constants but rather depend on the worms and kinks presented in the current configuration.
It also should be noted that every procedure has the inverse one. The probabilities of such procedures must be chosen according to those of the direct ones.

$\textbf{(iii)}$ the site $i$ and times $\tau_1$ and $\tau_2$ are chosen according to the procedure.
The site $i$ is chosen directly in case of the worm-dependent procedure and randomly otherwise. Moments of time are calculated in accordance to the probability density calculated for each case.

$\textbf{(iv)}$the probability $W$ of accepting new configuration is calculated. If $W>R$, where $R$ is a random number from the interval $[0,1]$, the new configuration is accepted.
If the updating procedure is interrupted due to the impossibility of the updates, it corresponds to the case $W=0$. However, such interruptions are part of the statistics and they occur in accordance with the detail balance.

$\textbf{(v)}$ in case the system has no worms regardless of the accepting of the new configuration, the statistics is supplemented by the new data and the procedure goes back to the step (ii), otherwise the procedure goes to the step (ii) without the data supplementing.

Furthermore, by the fact that the CPU time depends linearly on the lattice size and the average sign remains large enough, which helps to keep errors in acceptable limits, the simulation can be made at a relatively large lattice size. All numerical results were obtained for a $20\times20$ lattice cluster with periodic boundary condition. However, this size is not enough to ensure that the finite-size effects have no significant effects on the result. To make these effects negligible the size of the lattice cluster should be extended to at least the $30\times30$ one.

\section{Results}

\begin{figure}
\begin{minipage}[h]{1\linewidth}
\center{\includegraphics[width=1\linewidth]{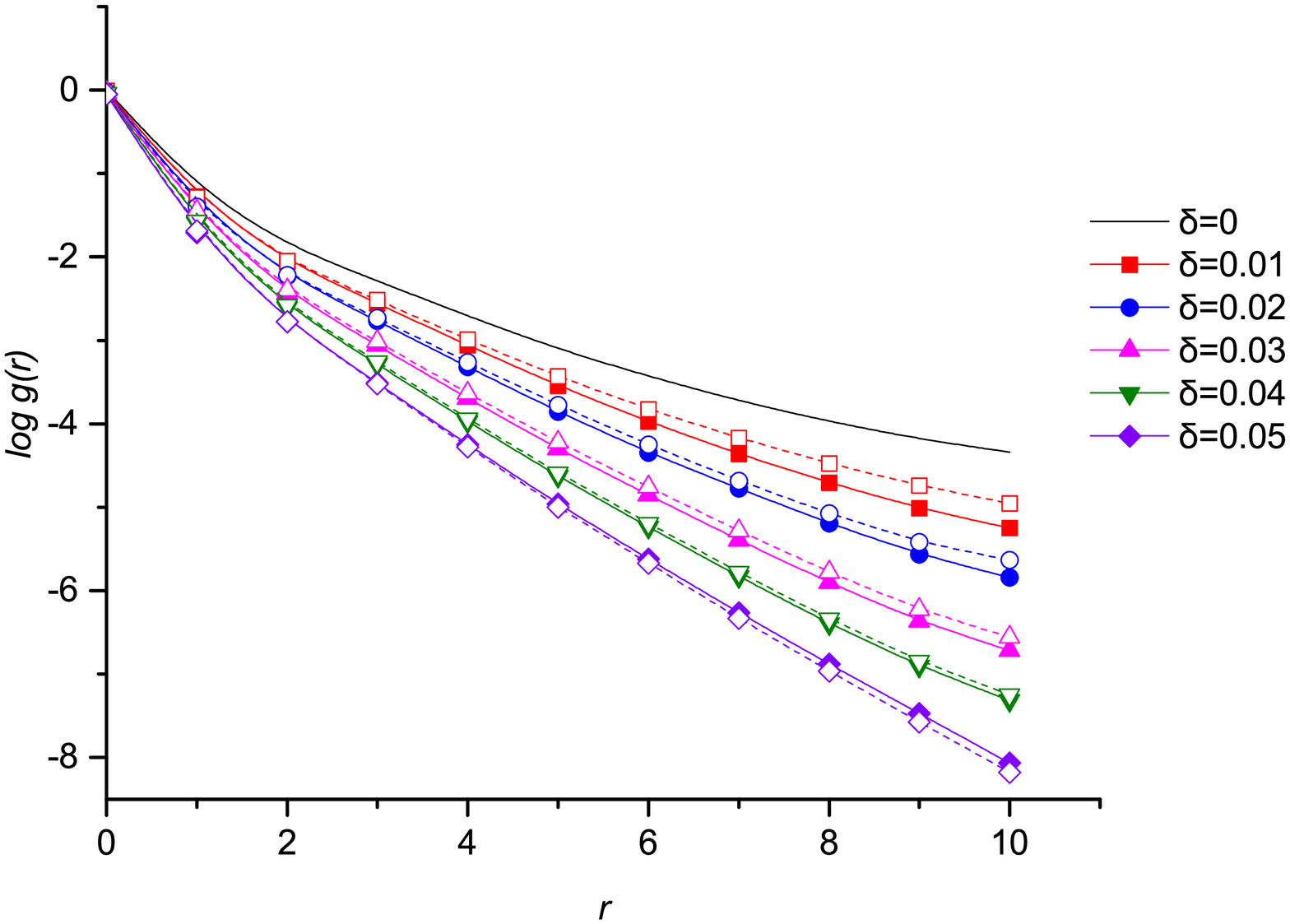}} a) \\
\end{minipage}
\hfill
\begin{minipage}[h]{1\linewidth}
\center{\includegraphics[width=1\linewidth]{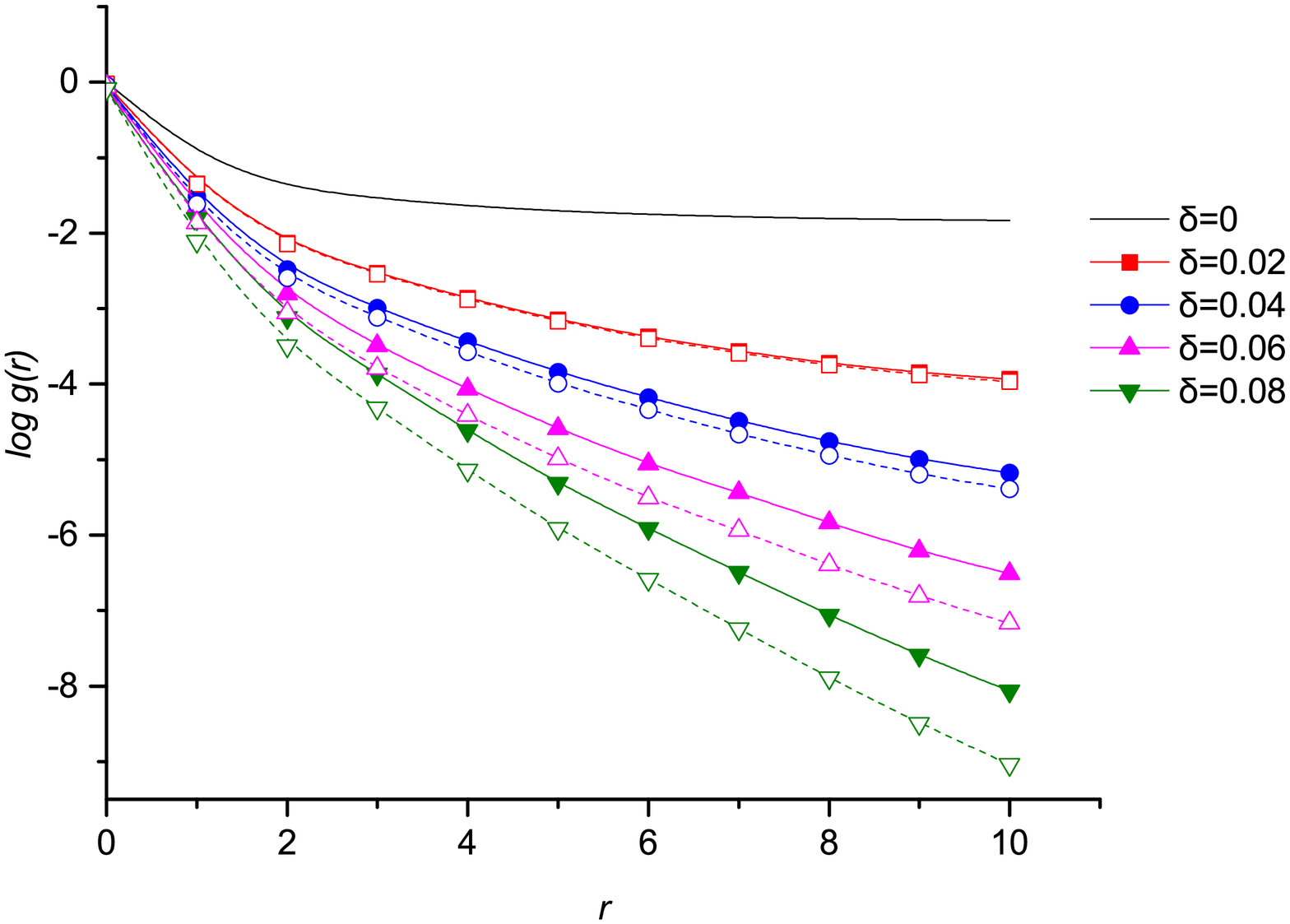}} b) \\
\end{minipage}
\caption{Panel (a) and (b) show $\log g(r)$ for $J=0.2t$ (a) and $J=0.4t$ (b) with $T=0.1t$. Solid (dashed) lines show results obtained for $t'=t''=0 (t'=-0.27t, t''=0.2t)$, respectively.}
\label{ris:logg(r)}
\end{figure}

\begin{figure}
\begin{minipage}[h]{1\linewidth}
\center{\includegraphics[width=1\linewidth]{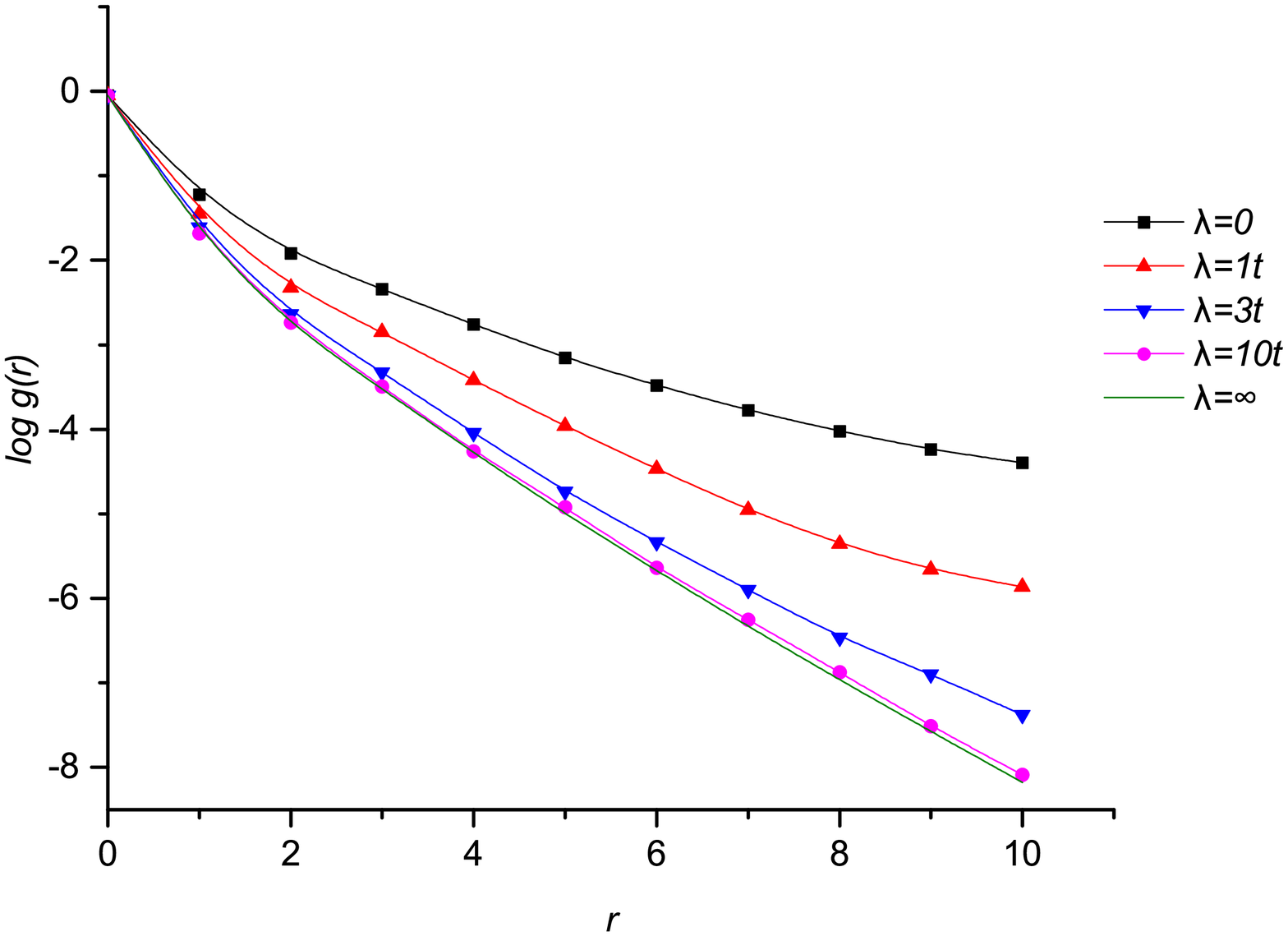}} a) \\
\end{minipage}
\hfill
\begin{minipage}[h]{1\linewidth}
\center{\includegraphics[width=1\linewidth]{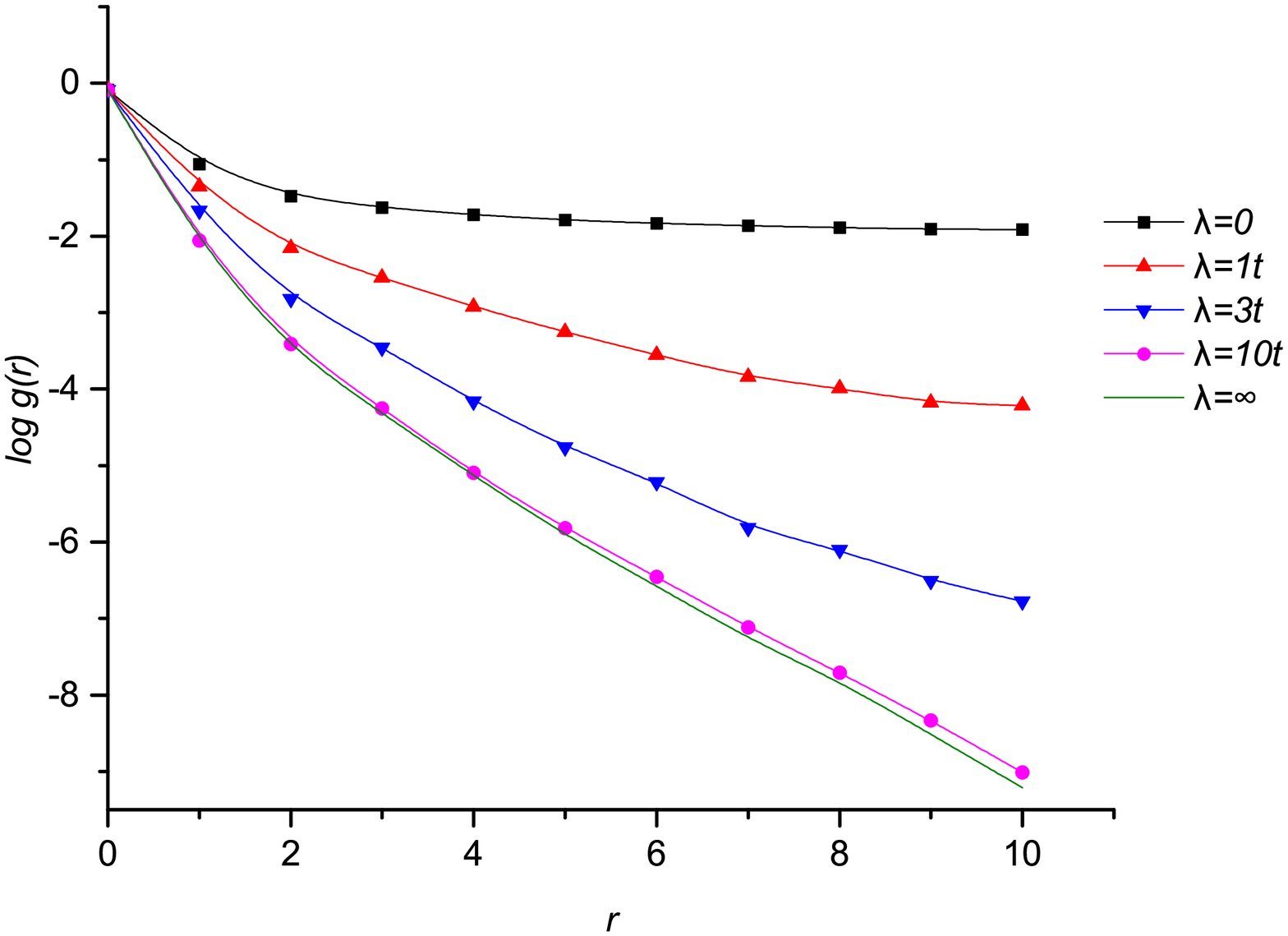}} b) \\
\end{minipage}
\caption{Curves (a) and (b) show $\log g(r)$ for $J=0.2t$, $\delta=0.05$ (a) and $J=0.4t$, $\delta=0.08$ (b) with $T=0.1t$ and $t'=-0.27t, t''=0.2t$.}
\label{ris:lambda}
\end{figure}

To estimate the dependence of AF order from the doping level we compute the  spin-spin correlation function $g(r)$
for the physical electron operators.
This is calculated with $\sum_i{S_i^z}=\sum_i{s_i^z}=0$ and a fixed number of dopons, $\delta$:
\begin{eqnarray}
\begin{gathered}
g(r) = 4\Delta^{-1}(r)\sum_{ij}\sum_{pq}\langle p|S^z_i+s^z_i|p\rangle \langle q|S^z_j+s^z_j|q\rangle\times\\ \times \langle X^{pp}_i X^{qq}_j\rangle e^{i \mathbf{K}\cdot (\mathbf{R}_i-\mathbf{R}_j)}\bar{\delta}(r-|\mathbf{R}_i-\mathbf{R}_j|),
\label{06}
\end{gathered}
\end{eqnarray}
where $\mathbf{K}=(\pi,\pi)$, $\mathbf{R}_i$ is radius-vector of the site $i$, $\Delta(r)=\sum_{ij} \bar{\delta}(r-|\mathbf{R}_i-\mathbf{R}_j|)$ and
\begin{equation}
\bar{\delta}(x)=
    \begin{cases}
    1 &\text{if $|x|\le 0.5a$,}\\
    0 &\text{otherwise,}
    \end{cases}
\end{equation}
with $a$ being the lattice constant and $\langle...\rangle$ means an average over the spin configurations generated in the QMC run.
In all the figures showing $g(r)$ we use logarithmic scale for the vertical axis. Therefore, for the LRO, QLRO and SRO,
the $g(r)$ should be represented asymptotically by a constant, a logarithmic function,
and a straight line, respectively.

Fig.\ref{ris:logg(r)} displays the electron spin-spin correlators for the different doping levels.
The critical hole concentration varies from
around $\delta_c=0.05$ at $J=0.2t$ to $\delta_c=0.08$ at $J=0.4t$.
Due to a finite lattice size as well as a finite temperature a true long-range AF order
manifest itself as a QLRO even at a very small doping. The suppression of the true LRO
corresponds to the destruction of the QLRO due to the emergence of the short-range AF correlations.
The obtained values of the critical hole concentrations do not necessarily coincide with the true ones
to be computed at zero temperature in the thermodynamic limit.
However their magnitudes are reasonably small.

In Fig.\ref{ris:lambda} we report the spin-spin correlators, $\log g(r)$, for $J=0.2t$ and $J=0.4t$
at $\delta_c=0.05$ and $\delta_c=0.08$ for different values of $\lambda$, respectively.
It is clearly seen that the QLRO is restored as $\lambda$ decreases.
The local NDO constraint
plays the dominating role in the destruction of the long-range AF state.
At $\lambda>10t$,  the spin-spin correlation functions
become almost identical to each other.
This indicates that
finite but large enough values of $\lambda$ already provide a reliable description of the existing strong correlations.
In this limit, the high and low energy itinerant fermions cannot be separated out and this is another
manifestation of the duality of the lattice electron nature.

It should be noted that the physical meanings of the Kondo coupling $\lambda$ within the conventional
phenomenological spin-fermion model \cite{chub} and in our Eq.(\ref{04})
are completely different.
In the former case, it represents a SDW gap that can evolve from small to large values.
In our theory, the only meaningful value of the Kondo coupling is that of $\lambda\gg t$
to take proper care of the NDO constraint.

\section{Conclusion}

To conclude,  we investigate the spin-spin correlation functions in the underdoped $t-J$ model
numerically by employing quantum Monte Carlo simulations on finite clusters.
Our main conclusion is that it is the local NDO constraint that is behind the rapid suppression of the AF QLRO
at surprisingly small doping level. In contrast, any mean-field global treatment of the local NDO results in unphysically large
values of the critical hole concentration.

The itinerant-localized duality of the lattice electrons offers the
following explanation of the rapid destruction of the magnetic order
by strong correlations.
The localized individual lattice spins become less correlated
with each other due to the competition between the AF correlations (the characteristic energy scale $\sim J$) and the Kondo screening ($\sim \lambda$) of the local spin moments
by the conduction dopons. The screening breaks the AF bonds.
In case a double occupancy is allowed, this breaking is not very efficient, as
it is  induced by the small (in this regime) spin-dopon interaction $\lambda$.

As $\lambda$ increases, the screening becomes more effective.
Since $1/J\gg 1/t$, the hole dynamics is much faster than the spin one. The broken AF bonds recover themselves
at a much slower rate than the breaking occurs.
As a result,
even a small amount of fast moving dopons (holes) turns out to be, at a large enough $\lambda$,  sufficient
to completely destroy the AF LRO.

A further possible application of the present approach
might be that to theoretically explore an experimentally observed instability
towards a formation of a charge order in the pseudogap phase at $\delta\approx 0.1$
There is a strong evidence that the observed charge order is due to strong electron correlations \cite{neto}.
The spin-dopon representation of the $t-J$ model provides a natural framework to address this problem.
By varying $\lambda$, we would be able to vary the strength of the correlations to explicitly explore
the impact of the NDO on the charge order formation.
This is already in progress and results will be presented elsewhere.

\section{Appendix}

For the $1d$ $t-J$ model, the two leading terms of the ground-state energy expansion  in powers of $J/t\ll 1$ are known explicitly.
In the present Appendix, we show that the spin-dopon model (\ref{2.7}) produces exactly the same result.

As the sign of $t$ is
irrelevant,  we can fix the Hamiltonian (\ref{2.7}) in $1d$ to take the form
\begin{eqnarray}
H_{t-J}=H_{J=0}+H_{int},
\label{a2}\end{eqnarray}
where
\begin{eqnarray}
H_{J=0}&=:&H_0
=-2t\sum_{ij\sigma}{d}_{i\sigma}^{\dagger} {d}_{j\sigma}\nonumber\\
&+&\frac{3\lambda}{4}\sum_{i\sigma}{d}_{i\sigma}^{\dagger} {d}_{i\sigma} +\lambda
\sum_{i}\vec{S_i} \cdot \vec{s_i},\quad t> 0,
\label{a3}\end{eqnarray}
and
\begin{equation}
H_{int}=J\sum_{ij}\vec S_i\vec S_j(1-n^d_i)(1-n^d_j).
\label{a4}\end{equation}

The limit $\lambda\to \infty$
reduces the local Hilbert space to that comprising  a lattice spin-up state $|\uparrow \rangle_i=|\uparrow 0\rangle_i$,
a spin-down state
$|\uparrow \rangle_i =|\uparrow 0\rangle_i$ and  a vacancy state $|0\rangle_i=\frac{|\uparrow\downarrow\rangle_i-
|\downarrow\uparrow\rangle_i}{\sqrt{2}}$.
We define the basis of the one-vacancy states as
$$|i,\{\sigma\}\rangle=|\sigma_1\sigma_2...0_i...\sigma_N\rangle,$$
where $\sigma_k=\uparrow\downarrow$ and $\{\sigma\}$ is a multi-index describing an arbitrary set
of the lattice spins.
The vacancy state $|0\rangle_i$ is a total spin singlet defined above.

The ground state at $J=0$ is degenerate with respect to spin. We can therefore  choose
a FM spin configuration. An arbitrary one-hole state is then given by
\begin{equation}
|\Phi\rangle=\sum_i\phi_i|i,\{\uparrow\}\rangle.
\label{a6}\end{equation}
The energy of such a state is given by
\begin{equation}
\sum_{ij}\langle \Phi|H_{J=0}|\Phi\rangle=-\sum_{ij}t_{ij}\bar\phi_j\phi_i.
\label{a7}\end{equation}
The corresponding Schrodinger equation reads
\begin{equation}
\sum_{j}(t_{ij}-E\delta_{ij})\phi_j=0.
\label{a8}\end{equation}
The lowest-energy solution for the nearest-neighbour (nn) interaction reads
$$\phi_j=1/\sqrt{N_s},\quad E_0=-2t,$$
with $N_s$ being a total number of the lattice sites.

To consider a state with $N$ holes one should generalize Eq.(\ref{a6}) to include $N$ fermionic (hole) states:
$$|\Phi_{N}\rangle=\sum_{i_1,i_2,...,i_N}\phi_{i_1,i_2,...,i_N}|i_1,i_2,...,i_N\{\uparrow\}\rangle,$$
where the function $\phi_{i_1,i_2,...,i_N}$ is antisymmetric with respect to the index permutation.
A corresponding Schrodinger $N$-particle equation can be then written out explicitly.
Alternatively, one can quantize Eq.(\ref{a8}) with exactly the same effect. Namely,
the $c$-valued amplitudes $\phi_i$ are replaced by the fermion operators
$$\phi_i\to \hat\phi_i=:f_i,\quad [f_i^+,f_j]_{+}=\delta_{ij}.$$
The $N$-hole generalization of the Hamiltonian $H_0$ then reads
\begin{equation}
H_0=-\sum_{ij}t_{ij}
f^+_if_j,\quad \sum_i f_i^{\dagger}f_i=N.
\label{a9}\end{equation}
This Hamiltonian describes spinless fermions hopping in a $1d$ lattice.
In case of the nn interaction,
the ground-state energy becomes
$$E_0=-\frac{2t}{\pi}\sin(\pi\delta),\quad \delta=\frac{N}{N_s}=\frac{1}{N_s}\langle \sum_i f^+_if_i\rangle_{H_0}.$$

The spin degeneracy is lifted by the effective spin-spin interaction:
\begin{equation}
H_{t-J}^{gr}=-\frac{2t}{\pi}\sin(\pi\delta)+J_{eff}\sum_{ij}\vec S_i\vec S_j +{\cal O}(J^2), \quad J\to 0.
\label{a10}\end{equation}
We have
$$J_{eff}=J\langle(1-f^{\dagger}_{i}f_i)(1-f^{\dagger}_{j}f_j)\rangle_{H_0}$$
$$=J((1-\delta)^2-\frac{\sin^2\pi(1-\delta)}{\pi^2}).$$
In terms of the electron density $n_e=1-\delta$,
Eq.(\ref{a10}) becomes
\begin{eqnarray}
H_{t-J}^{gr}&=&-\frac{2t}{\pi}\sin(\pi n_e)
\nonumber\\
&+& J(n_e^2-\frac{\sin^2\pi n_e}{\pi^2})\sum_{ij}\vec S_i\vec S_j +{\cal O}(J^2),
\label{final1}\end{eqnarray}
which agrees with the Bethe-ansatz result obtained for the canonical $t-J$ model given by Eq.(\ref{1.1})\cite{shiba}.

\end{document}